\documentclass{jpp}

\usepackage{amsmath}             % advanced math
\usepackage{amssymb}             % additional math symbols
\usepackage{graphicx}            % figures
\usepackage{hyperref}            % hyperlinks
\usepackage{subcaption}          % captioned subfigures

\usepackage{soul}                % underlining with \ul{}
\usepackage{xcolor}              % coloring

\newcommand{\vect}[1]{\boldsymbol{#1}}

\newcommand{\pd}{\partial}

\newcommand{\cross}{\times}
\newcommand{\grad}{\bnabla}

\renewcommand{\div}{\bnabla \bcdot}

\shorttitle{Plasma sheet thinning due to loss of magnetotail plasma}
\shortauthor{R.~Tretler, T.~Tatsuno and K.~Hosokawa}

\title{Plasma sheet thinning due to loss of near-Earth magnetotail plasma}

\author{
   Rudolf~Tretler\aff{1} \corresp{\email{rtretler@protonmail.com}},
   Tomo~Tatsuno\aff{2},
   \and Keisuke~Hosokawa\aff{2}
}

\affiliation{
   \aff{1}Department of Communication Engineering and Informatics \\
   \aff{2}Department of Computer and Network Engineering \\
   University of Electro-Communications, Tokyo 182-8585, Japan
}

\begin{document}

\maketitle

\begin{abstract}
A one-dimensional model for thinning of the Earth's plasma sheet
[J.~K.~Chao et al., Planet.~Space Sci.~\textbf{25}, 703 (1977)]
according to the Current Disruption (CD) model of auroral breakup
is extended to two dimensions.
A rarefaction wave, which is a signature component of the CD model,
is generated with an initial disturbance.
In the 1D gas model, the rarefaction wave propagates tailward
at sound velocity and is assumed to cause thinning.
Extending to a 2D gas model of a simplified plasma sheet configuration,
the rarefaction wave is weakened, and the thinning ceases to propagate.
Extending further to a 2D plasma model
by adding magnetic field into the lobes,
the rarefaction wave is quickly lost
in the plasma sheet recompression,
but the plasma sheet thinning is still present
and propagates independently at a slower velocity than a 1D model suggests.
This shows that the dynamics of plasma sheet thinning may be dominated
by sheet--lobe interactions that are absent from the 1D model
and may not support the behaviour assumed by the CD model.
\end{abstract}


%%%%% MAIN TEXT %%%%%

%%%%%%%%%%%%%%%%%%%%%%%%%%%%%%%%%%%%%%%%%%%%%%%%%%%%%%%%%%%%%%%%%%%%%%%%%%%%%%%%
%%%%%%%%%%%%%%%%%%%%%%%%%%%%%%%%%%%%%%%%%%%%%%%%%%%%%%%%%%%%%%%%%%%%%%%%%%%%%%%%

\section{Introduction}
\label{sec:introduction}

The mechanisms behind auroral breakup,
a sudden increase of auroral strength during
the magnetospheric substorm~{\citep{akasofu1968book-magnetospheric-substorms}},
are not yet entirely understood.
While it is known~{\citep{schindler1975plasma}} that the three main events in this process are
a) magnetotail reconnection,
b) cross tail current reduction and
c) auroral breakup,
their exact order has not been conclusively determined.
There are two main competing models,
the Near-Earth Neutral Line (NENL) model
and the Current Disruption (CD) model.

In the Near-Earth Neutral Line model~{\citep{baker1996neutral}},
a reconnection of the magnetic field lines in the magnetotail
creates jets of plasma that flow earthward and tailward.
The earthward jet causes a decrease in the cross tail current,
and then enters the high-latitude atmosphere,
where it causes the auroral breakup.

\begin{figure}
   \centerline{
      \includegraphics[width=0.45\textwidth]{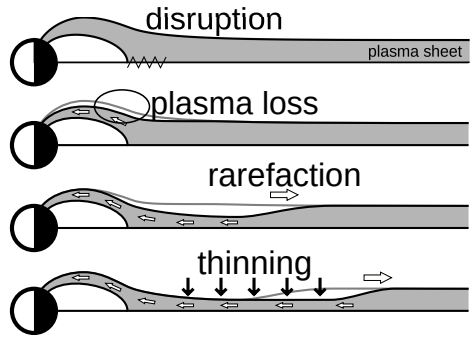}
   }
   \caption{A rarefaction wave causing plasma sheet thinning.
     Image adapted from {\citet{chao_plasma}}.}
   \label{fig:plasma-sheet-thinning}
\end{figure}

In the Current Disruption model~{\citep{lui1977search}},
a current disruption instability reduces the cross tail current,
breaking the balance of the near-Earth magnetotail plasma,
which enters the high-latitude atmosphere and causes the auroral breakup.
The magnetotail plasma loss induces
a rarefaction wave in the plasma sheet (Fig.~{\ref{fig:plasma-sheet-thinning}}).
The tailward propagation of the rarefaction wave causes
a reduction of plasma sheet thickness,
eventually leading to magnetotail reconnection.

About a decade ago, the THEMIS mission~{\citep{angelopoulos2008themis}} tried to determine
which of these two models is the one triggering the auroral substorms.
Satellite part of the THEMIS mission is composed of five identical satellite probes
distributed and coordinated in the tail of the magnetosphere,
which enables us to observe the disturbances in the magnetotail leading to the auroral substorm.
To support the satellite observations in space,
a number of all-sky cameras were deployed in North America to observe the signature
of auroral breakup at the magnetic footprints of the satellites~{\citep{mende2008themis}}.
By using the data obtained by the THEMIS mission, {\citet{angelopoulos2008reconnection}}
demonstrated an event in which reconnection took place at $-20 R_\textrm{e}$ a few minutes earlier
than the signature of current disruption at $\sim -10 R_\textrm{e}$.
This observation supports the NENL model, that is,
that auroral substorms are initiated by reconnection in the magnetotail.
Later, however, {\citet{lui2009comment}} claimed that multi-satellite observations
during the same interval can be interpreted based on the CD paradigm.
Thus, it is still controversial which of these two models better explains the development
of the magnetotail disturbances before auroral breakups,
and can be regarded as the dominant triggering mechanism of substorms.

In this paper, we consider the CD model.
{\citet{chao_plasma}} have approximated the initial disturbance
that causes the rarefaction wave with
an imaginary piston on the near-Earth side of the plasma sheet;
earthward movement of the piston generates the rarefaction wave.
A simplified model of a piston-bounded 1D gas tube was used
to approximate the rarefaction wave in the weakly magnetized neutral sheet plasma,
and the result extrapolated to the full plasma sheet.

We extend the model to a 2D vertical cross-section of the plasma sheet,
including the north and south magnetic lobes.
This allows us to take into account the influence of
the strongly magnetized lobe plasma,
as well as any dynamics that result from the interaction of
plasmas as radically different as sheet and lobe plasma are.
We show that, although there is a small drop in pressure,
the rarefaction wave, which is supposed to be a signature of the CD model, is not noticeable.
Furthermore, the propagation velocity of the thinning front shows
a strong dependence on lobe conditions.

This paper is organized as follows.
In Section~{\ref{sec:simulation}},
we introduce the magnetohydrodynamic (MHD) equations,
the physical model of the plasma sheet, and the numerical scheme.
In Sections~{\ref{sec:results-gas}} and {\ref{sec:results-plasma}},
simulation results for the plasma sheet modelled with,
respectively, gas equations and MHD equations are presented and discussed.
In Section~{\ref{sec:comparison}},
we compare the results of the 1D and 2D models.
Finally, concluding remarks are given in Section~{\ref{sec:conclusion}}.

%%%%%%%%%%%%%%%%%%%%%%%%%%%%%%%%%%%%%%%%%%%%%%%%%%%%%%%%%%%%%%%%%%%%%%%%%%%%%%%%
%%%%%%%%%%%%%%%%%%%%%%%%%%%%%%%%%%%%%%%%%%%%%%%%%%%%%%%%%%%%%%%%%%%%%%%%%%%%%%%%

\section{Simulation setup}
\label{sec:simulation}

%%%%%%%%%%%%%%%%%%%%%%%%%%%%%%%%%%%%%%%%%%%%%%%%%%%%%%%%%%%%%%%%%%%%%%%%%%%%%%%%
%%%%%%%%%%%%%%%%%%%%%%%%%%%%%%%%%%%%%%%%%%%%%%%%%%%%%%%%%%%%%%%%%%%%%%%%%%%%%%%%

\subsection{MHD equations}
\label{sec:mhd-equations}

We use the normalized MHD equations in their formulation as
a system of conservation laws~{\citep{ryu1995multidimensional}}.
System of conservation laws in 2D is
\begin{equation}
   \label{eq:conservation-law-2d}
   \frac{\pd \vect{U}}{\pd t}
   + \frac{\pd }{\pd x}F(\vect{U})
   + \frac{\pd }{\pd y}G(\vect{U})
   = 0,
\end{equation}
where $\vect{U}$ is a vector of conserved variables and
$F(\vect{U})$ and $G(\vect{U})$ are respectively fluxes in $x$ and $y$ directions.
For the ideal MHD system, the conserved variables are
\begin{equation}
   \label{eq:variable-vector}
   \vect{U} = \left(
        \rho
      , \rho u
      , \rho v
      , \rho w
      , B_x
      , B_y
      , B_z
      , e
   \right)^T,
\end{equation}
where $\rho$ is the density,
$\vect{u} = ( u, v, w )$ is the velocity vector,
$\vect{B} = ( B_x, B_y, B_z )$ is the magnetic field vector,
and $e$ is the total energy.
The fluxes in~{\eqref{eq:conservation-law-2d}} are
\begin{equation}
   \label{eq:flux-vector-f}
   F(\vect{U}) = \left[ \begin{array}{l}
      \rho u \\
      \rho u u - B_x B_x + p_{\mathrm{total}} \\
      \rho v u - B_x B_y \\
      \rho w u - B_x B_z \\
      0 \\
      B_y u - B_x v \\
      B_z u - B_x w \\
      u(e + p_{\mathrm{total}}) - B_x(\vect{u} \bcdot \vect{B})
   \end{array} \right],
\end{equation}
\begin{equation}
   \label{eq:flux-vector-g}
   G(\vect{U}) = \left[ \begin{array}{l}
      \rho v \\
      \rho u v - B_y B_x \\
      \rho v v - B_y B_y + p_{\mathrm{total}} \\
      \rho w v - B_y B_z \\
      B_x v - B_y u \\
      0 \\
      B_z v - B_y w \\
      v(e + p_{\mathrm{total}}) - B_y(\vect{u} \bcdot \vect{B})
   \end{array} \right],
\end{equation}
where total pressure $p_{\mathrm{total}}$ is
\begin{equation}
   \label{eq:ptotal}
   p_{\mathrm{total}}
   = p
   + \frac{1}{2} \vect{B} \bcdot \vect{B}
\end{equation}
with plasma kinetic pressure $p$ defined as
\begin{equation}
   \label{eq:pressure}
   p
   = (\gamma-1)
   \left(
      e
      - \frac{1}{2} \rho \vect{u} \bcdot \vect{u}
      - \frac{1}{2} \vect{B} \bcdot \vect{B}
   \right),
\end{equation}
where $\gamma$ is the ratio of specific heats, taken to be $5/3$.

%%%%%%%%%%%%%%%%%%%%%%%%%%%%%%%%%%%%%%%%%%%%%%%%%%%%%%%%%%%%%%%%%%%%%%%%%%%%%%%%
%%%%%%%%%%%%%%%%%%%%%%%%%%%%%%%%%%%%%%%%%%%%%%%%%%%%%%%%%%%%%%%%%%%%%%%%%%%%%%%%

\subsection{Plasma sheet model}
\label{sec:plasma-sheet-model}

\begin{figure}
   \centerline{
      \includegraphics[width=0.60\textwidth]{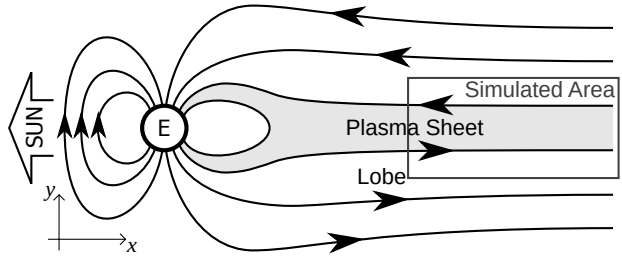}
   }
   \caption{Rough structure of the Earth's magnetosphere, with the simulated area marked.}
   \label{fig:plasma-sheet-structure}
\end{figure}

\begin{figure}
   \centerline{
      \includegraphics[width=0.60\textwidth]{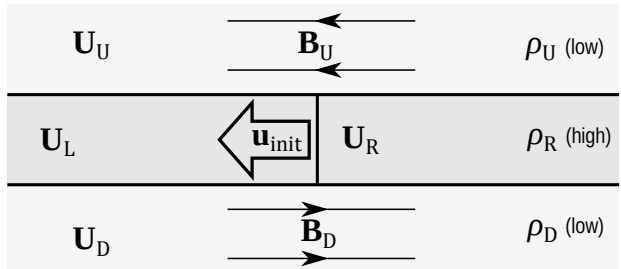}
   }
   \caption{Initial configuration of the simulated area of the plasma sheet.}
   \label{fig:plasma-sheet-init}
\end{figure}

For the simulation, we take the relatively flat area of the plasma sheet,
where the magnetic field lines are approximately parallel (Fig.~{\ref{fig:plasma-sheet-structure}}),
and look at the $x$-$y$ cross section,
where the $x$ axis points from the Earth tailward, and $y$ axis points north.

In the neutral sheet is a weakly-magnetized, high-density plasma, $\vect{U}_\mathrm{R}$
(where ``R'' stands for ``Right'';
$\vect{U}_\mathrm{L}$, ``Left'',
is the initial disturbance region and will be described later; see Fig.~{\ref{fig:plasma-sheet-init}}).
This plasma is sandwiched between the magnetic lobes,
with strongly-magnetized, antiparallel, low-density plasmas,
$\vect{U}_\mathrm{U}$ (northern lobe, ``Up'')
and $\vect{U}_\mathrm{D}$ (southern lobe, ``Down'').

The inner layer of the plasma sheet,
$\vect{U}_\mathrm{sheet} = \vect{U}_\mathrm{R,L}$,
was assumed by {\citet{chao_plasma}} to have a profile described by
$\vect{B}_\mathrm{sheet} = \vect{B}_\infty \tanh(y)$.
However, testing has shown that the results are almost identical
if the plasma sheet contains uniform plasma with no magnetic field.
Since the latter is more amenable to analysis,
we use a uniform plasma sheet as the initial condition for our simulations.

We assume that prior to the disturbance
the plasma sheet was in a steady-state configuration,
in which case sheet and lobes are separated by tangential discontinuities.
The Rankine-Hugoniot jump condition for a tangential discontinuity~{\citep{baumjohannplasma}} is
\begin{equation}
   \label{eq:rankine-hugoniot-condition}
   \left[ p_{\mathrm{total}} \right] = 0,
\end{equation}
where $[X]$ denotes the jump in $X$ when crossing the boundary.
With plasma sheet magnetic field $\vect{B}_{\mathrm{sheet}} = \vect{0}$,
and lobe magnetic field pointing in the $x$ direction,
$\vect{B}_\mathrm{U,D} = \vect{B}_{\mathrm{lobe}} = (\pm B_x, 0, 0)$, this condition becomes
\begin{equation}
   \label{eq:pressure-balance}
   p_{\mathrm{sheet}}
   = p_{\mathrm{lobe}} + \frac{1}{2}B_{x,\mathrm{lobe}}^2.
\end{equation}

We normalize the system so that $p_\mathrm{sheet} = 1.0$, $\rho_\mathrm{sheet} = 1.0$,
and the initial thickness of the plasma sheet is $h_\mathrm{sheet} = 1.0$.
The relationship between physical and normalized units,
as well as realistic values for sheet and lobe
obtained from satellite measurements~{\citep{baumjohannplasma,baumjohann1989average}},
are shown in table~{\ref{tab:normalization}}.
With the geometry fixed and physical quantities normalized to sheet conditions,
we can fully describe the problem with only a handful of parameters.

As a first parameter we take the \emph{lobe plasma beta} $\beta_\mathrm{lobe}$,
where plasma beta is defined as $\beta = 2p / B^2$.

For the second parameter, we define the kinetic \emph{temperature ratio} $\tau$,
\begin{equation}
   \label{eq:temperature-ratio}
   \tau
   = \frac{T_\mathrm{i,sheet}}{T_\mathrm{i,lobe}}
   = \frac{p_\mathrm{sheet} / \rho_\mathrm{sheet}}{p_\mathrm{lobe} / \rho_\mathrm{lobe}}
\end{equation}
where the ion temperature $T_\mathrm{i}$ is defined through
\begin{align}
   \label{eq:temperature-to-pressure} p    &= n_\mathrm{i} k_\mathrm{B} T_\mathrm{i} \\
   \label{eq:particles-to-density}    \rho &= n_\mathrm{i} m_\mathrm{p},
\end{align}
where $n_\mathrm{i}$ is the ion number density,
$k_\mathrm{B}$ is the Boltzmann constant,
and $m_\mathrm{p}$ is the proton mass.

These two parameters, plasma beta $\beta_\mathrm{lobe}$ and temperature ratio $\tau$,
are sufficient to define the steady-state initial condition of the normalized plasma sheet.

For the initial disturbance, the piston model used by {\citet{chao_plasma}} is replaced
with a simple earthward plasma flow which will induce the rarefaction wave.
The flow is created by assigning an initial velocity
$\vect{u}_\mathrm{init} = (u_\mathrm{init},0,0)$ to
the plasma $\vect{U}_\mathrm{L}$ on the Earth side of
the plasma sheet (Fig.~{\ref{fig:plasma-sheet-init}}).

The velocity magnitude $u_\mathrm{init}$, which indicates the strength of the disturbance,
is the third and final parameter needed to unambiguously define the plasma sheet problem.

Finally, we also define the sound velocity $c_\mathrm{s}$
and the Alfv\'en velocity $c_\mathrm{A}$ as
\begin{equation}
   \label{eq:sound-velocity}
   c_\mathrm{s} = \sqrt{\frac{\gamma p}{\rho}}, \quad
   c_\mathrm{A} = \sqrt{\frac{B^2}{\rho}}.
\end{equation}

\begin{table}
   \centering
   \begin{tabular}{c|c|c|c|c|c|c|}
      & time,
      & length,
      & velocity,
      & density,
      & pressure,
      & mag.~field, \\
      & $t$ (s)
      & $l$ (km)
      & $u$ (km/s)
      & $\rho$ (kg/m$^3$)
      & $p$ (nPa)
      & $B$ (nT) \\
      $1.0$ normalized units
      & $29.7$ & $1.91\cross10^5$ & $642$ & $8.35\cross10^{-22}$ & $0.345$ & $20.8$ \\
      realistic, sheet
      & --- & --- & --- & $8.35\cross10^{-22}$ & $0.345$ & $10$ \\
      realistic, lobe
      & --- & --- & --- & $1.67\cross10^{-23}$ & $0.00069$ & $30$ \\
   \end{tabular}
   \caption{Units are normalized with respect to the plasma sheet.
            The first row of data shows the relationship between physical units and normalized units.
            The second and third rows show the realistic values for plasma sheet and magnetic lobe,
            obtained from satellite measurements~{\citep{baumjohannplasma,baumjohann1989average}}.}
   \label{tab:normalization}
\end{table}

%%%%%%%%%%%%%%%%%%%%%%%%%%%%%%%%%%%%%%%%%%%%%%%%%%%%%%%%%%%%%%%%%%%%%%%%%%%%%%%%
%%%%%%%%%%%%%%%%%%%%%%%%%%%%%%%%%%%%%%%%%%%%%%%%%%%%%%%%%%%%%%%%%%%%%%%%%%%%%%%%

\subsection{Numerical scheme}
\label{sec:simulation-scheme}

Since the problem setup introduced in the previous section contains discontinuities,
and the solution is likely to contain shocks and rarefaction waves,
we require a numerical scheme that can safely handle them.
Preliminary testing has shown that, for our purposes,
the second-order Essentially Non-Oscillatory (ENO) scheme with
Lax-Friedrichs (LF) flux splitting~{\citep{harten_eno,shu_implement}}
has a good balance between accuracy,
computation speed, and complexity of implementation.

When the ENO scheme is applied to a system of equations,
we use the characteristic decomposition to transform the problem
into a set of mutually independent equations.
The characteristic decomposition requires the system to be hyperbolic,
i.e., to have a full set of eigenvalues and left and right eigenvectors of the system's Jacobians,
\begin{equation}
   \label{eq:jacobian}
   A(\vect{U})
   = \frac{\pd F}{\pd \vect{U}}, \quad
   B(\vect{U})
   = \frac{\pd G}{\pd \vect{U}}.
\end{equation}
However, if the MHD equations are expressed as a full 2D system of conservation laws,
we can see from {\eqref{eq:conservation-law-2d}} that
the fluxes $F$ and $G$ are guaranteed to contain at least one zero,
which means that their Jacobians are singular and the ENO scheme cannot be constructed.
To work around this problem, instead of solving the entire 2D system,
we split it up into collections of orthogonal 1D systems of conservation laws~{\citep{shu_implement}}.
This allows us to use the well-formed one-dimensional ENO-LF solver.

Using a one-dimensional solver on a two-dimensional MHD system solves one problem, but creates another.
As the $\div \vect{B} = 0$ condition is not explicitly enforced in the MHD equations,
the independent calculations in $x$ and $y$ direction are likely to introduce an error and the divergence becomes non-zero.
This error accumulates exponentially~{\citep{powell1994icase}}.
To remedy this issue, after every time step we conduct divergence cleaning by solving the Poisson equation
\begin{equation}
   \label{eq:divergence-cleaning-poisson}
   \nabla^2\phi + \div \vect{B} = 0
\end{equation}
with the SOR (Successive Over-Relaxation) method,
and calculating the corrected magnetic field~{\citep{ryu1995multidimensional}} with
\begin{equation}
   \label{eq:divergence-cleaning-correction}
   \vect{B}_\mathrm{corrected} = \vect{B} + \grad\phi.
\end{equation}

Finally, for time stepping we use
the optimal third-order TVD (Total Variation Diminishing) Runge-Kutta method~{\citep{gottliebshu1998tvdrk}}
with a variable time step $\Delta t$,
calculated so that the CFL number is lower than $0.1$.
The source code of the simulation program is available on GitHub~{\citep{web-github-mhd-simulation-code}}.

For the results presented in this paper, the simulation box length was $( L_x, L_y ) = ( 32.0, 6.0 )$ units,
with $-16 \leq x \leq 16$, $-3 \leq y \leq 3$.
Each unit length in $x$ and $y$ direction is divided into
$16$, $32$ and $64$ grid points, for a total of, respectively,
$( N_x, N_y ) = ( 512, 96 )$, $( 1024, 192 )$ and $(2048, 384)$ grid points.
The boundary conditions are Dirichlet at $x = -16.0$ (Earth) and $x = 16.0$ (tail),
Neumann at $y = -3.0$ (south) and $y = 3.0$ (north),
except for the magnetic field component perpendicular to the boundary,
which is calculated from $\div \vect{B} = 0$.
To reduce numerical artefacts,
the discontinuities in the initial conditions have been smeared over two additional grid points.

%%%%%%%%%%%%%%%%%%%%%%%%%%%%%%%%%%%%%%%%%%%%%%%%%%%%%%%%%%%%%%%%%%%%%%%%%%%%%%%%
%%%%%%%%%%%%%%%%%%%%%%%%%%%%%%%%%%%%%%%%%%%%%%%%%%%%%%%%%%%%%%%%%%%%%%%%%%%%%%%%

\section{Gas model, simulation and results}
\label{sec:results-gas}

To set up a baseline behaviour for comparison,
we first run the simulations without magnetic field,
essentially treating plasma as a gas.

%%%%%%%%%%%%%%%%%%%%%%%%%%%%%%%%%%%%%%%%%%%%%%%%%%%%%%%%%%%%%%%%%%%%%%%%%%%%%%%%
%%%%%%%%%%%%%%%%%%%%%%%%%%%%%%%%%%%%%%%%%%%%%%%%%%%%%%%%%%%%%%%%%%%%%%%%%%%%%%%%
\subsection{1D gas model}
\label{sec:results-gas-1d}

\begin{figure}
   \centerline{
      \hfill
      \begin{subfigure}[b]{0.45\textwidth}
         \includegraphics[width=\textwidth]{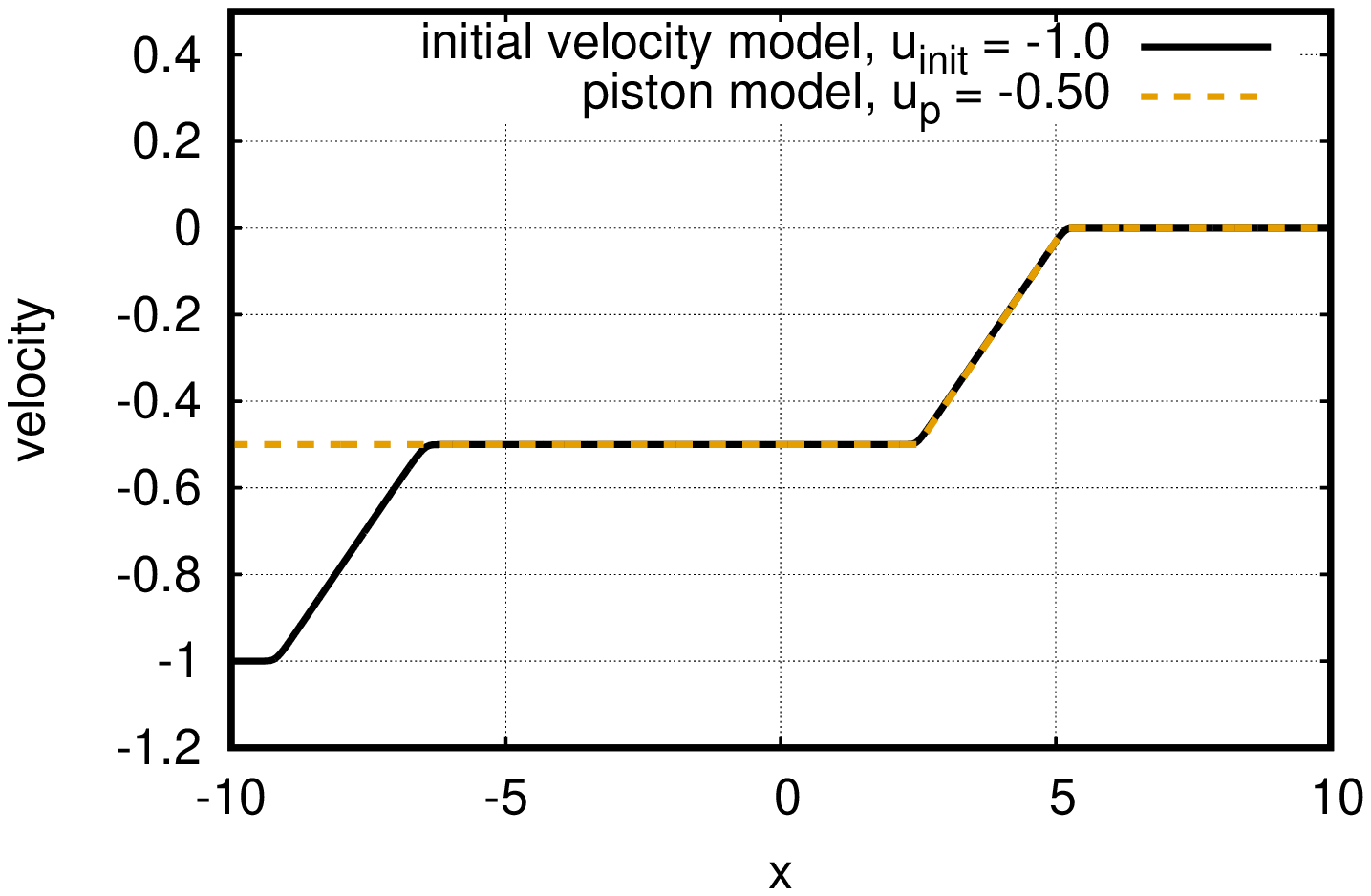}
         \caption{velocity at $t = 4.0$}
         \label{fig:result-1d-comp-v}
      \end{subfigure}
      \hfill
      \begin{subfigure}[b]{0.45\textwidth}
         \includegraphics[width=\textwidth]{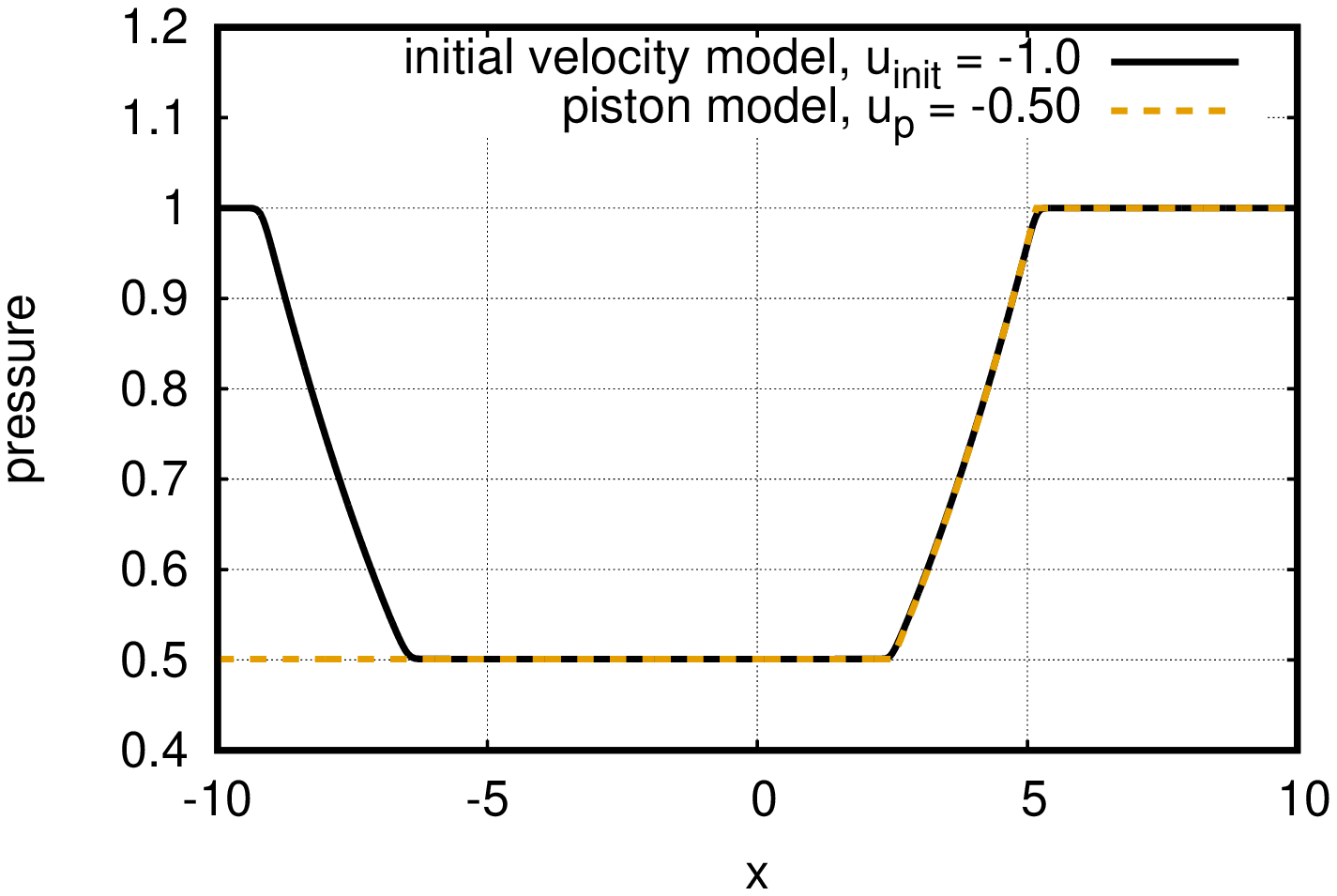}
         \caption{pressure at $t = 4.0$}
         \label{fig:result-1d-comp-p}
      \end{subfigure}
      \hfill
   }
   \caption{Results of the numerical simulation of the initial velocity model (solid line)
            compared to the exact solution of the equivalent piston model (dashed line).
            The plots shown are for~({\subref{fig:result-1d-comp-v}}) velocity
            and~({\subref{fig:result-1d-comp-p}}) pressure at time $t = 4.0$.}
   \label{fig:result-1d-comp}
\end{figure}

In the original 1D model, where the disturbance is generated with a piston~{\citep{chao_plasma}},
time evolution of plasma has an exact solution~{\citep{landau1987fluid}}.
If piston is moving at velocity $u_\mathrm{p}$, then for $0 < -u_\mathrm{p} < 2 c_\mathrm{s}/(\gamma-1)$,
plasma velocity $u(x,t)$, pressure $p(x,t)$, and density $\rho(x,t)$ are given by
\begin{align}
   u(x,t) &= \begin{cases}
      u_\mathrm{p} &\textrm{if } x < \left( c_\mathrm{s} + \frac{\gamma+1}{2}u_\mathrm{p} \right) t, \\
      0 &\textrm{if } x > c_\mathrm{s} t, \\
      \frac{2}{(\gamma+1)t} x - \frac{2 c_\mathrm{s}}{\gamma+1} & \textrm{otherwise},
   \end{cases} \\
   p(x,t) &= p_0 \left[ 1 - \frac{\gamma-1}{2} \frac{|u(x,t)|}{c_\mathrm{s}} \right] ^{2\gamma/(\gamma-1)}, \\
   \rho(x,t) &= \rho_0 \left[ 1 - \frac{\gamma-1}{2} \frac{|u(x,t)|}{c_\mathrm{s}} \right] ^{2/(\gamma-1)},
\end{align}
where $p_0, \rho_0$ are the initial pressure and density inside the plasma sheet.

In our model, the disturbance is generated with an initial velocity, $u_\mathrm{init}$.
If we perform a coordinate transformation so that the frame moves with the velocity $u_\mathrm{init}$,
we should obtain the exact same solution (though mirrored),
with the initial disturbance moving to the right with velocity $-u_\mathrm{init}$.
It follows that the solution is symmetrical around a point moving at half the initial velocity, $u_\mathrm{init}/2$.
Furthermore, the plasma at the point of symmetry should move at the same velocity as the point itself does.
We can conclude that the initial velocity model with an initial velocity $u_\mathrm{init}$ is equivalent to
the piston model with piston velocity $u_\mathrm{p} = u_\mathrm{init}/2$.

Figure~{\ref{fig:result-1d-comp}} shows the comparison of piston model in 1D theory~{\citep{chao_plasma}}
and the 1D simulation of the initial velocity model with a grid density of 128 points per unit length.
There is an excellent agreement between the results, confirming the equivalence of the two models.

%%%%%%%%%%%%%%%%%%%%%%%%%%%%%%%%%%%%%%%%%%%%%%%%%%%%%%%%%%%%%%%%%%%%%%%%%%%%%%%%
%%%%%%%%%%%%%%%%%%%%%%%%%%%%%%%%%%%%%%%%%%%%%%%%%%%%%%%%%%%%%%%%%%%%%%%%%%%%%%%%

\subsection{2D gas model}
\label{sec:results-gas-2d}

\begin{table}
   \centering
   \begin{tabular}{c||c|c||c|c|c|c||c|c|}
        Run
      & $\tau$
      & $u_\mathrm{init}$
      & $\rho_\mathrm{sheet}$
      & $p_\mathrm{sheet}$
      & $\rho_\mathrm{lobe}$
      & $p_\mathrm{lobe}$
      & $c_\mathrm{s,sheet}$
      & $c_\mathrm{s,lobe}$ \\
      A & $1.0$ & $-1.0$ & $1.0$ & $1.0$ & $1.0$ & $1.0$ & $1.29$ & $1.29$ \\
      B & $2.0$ & $-1.0$ & $1.0$ & $1.0$ & $2.0$ & $1.0$ & $1.29$ & $0.91$ \\
      C & $5.0$ & $-1.0$ & $1.0$ & $1.0$ & $5.0$ & $1.0$ & $1.29$ & $0.58$ \\
   \end{tabular}
   \caption{Initial conditions for the 2D unmagnetized plasma simulations.}
   \label{tab:initial-gas}
\end{table}

\begin{figure}
   \centerline{
      \hfill
      \begin{subfigure}[b]{0.45\textwidth}
         \includegraphics[width=\textwidth]{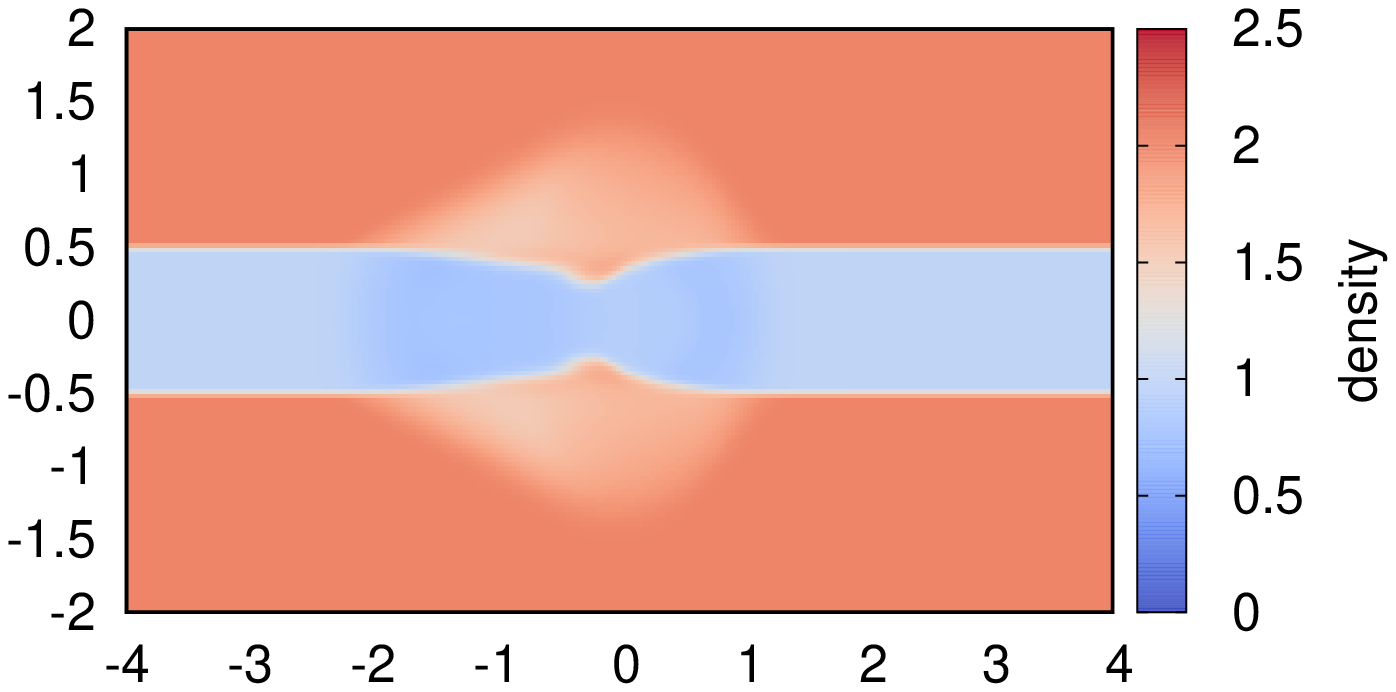}
         \caption{density at $t = 1.0$}
         \label{fig:result-plot-gas-d-t1-0}
      \end{subfigure}
      \hfill
      \begin{subfigure}[b]{0.45\textwidth}
         \includegraphics[width=\textwidth]{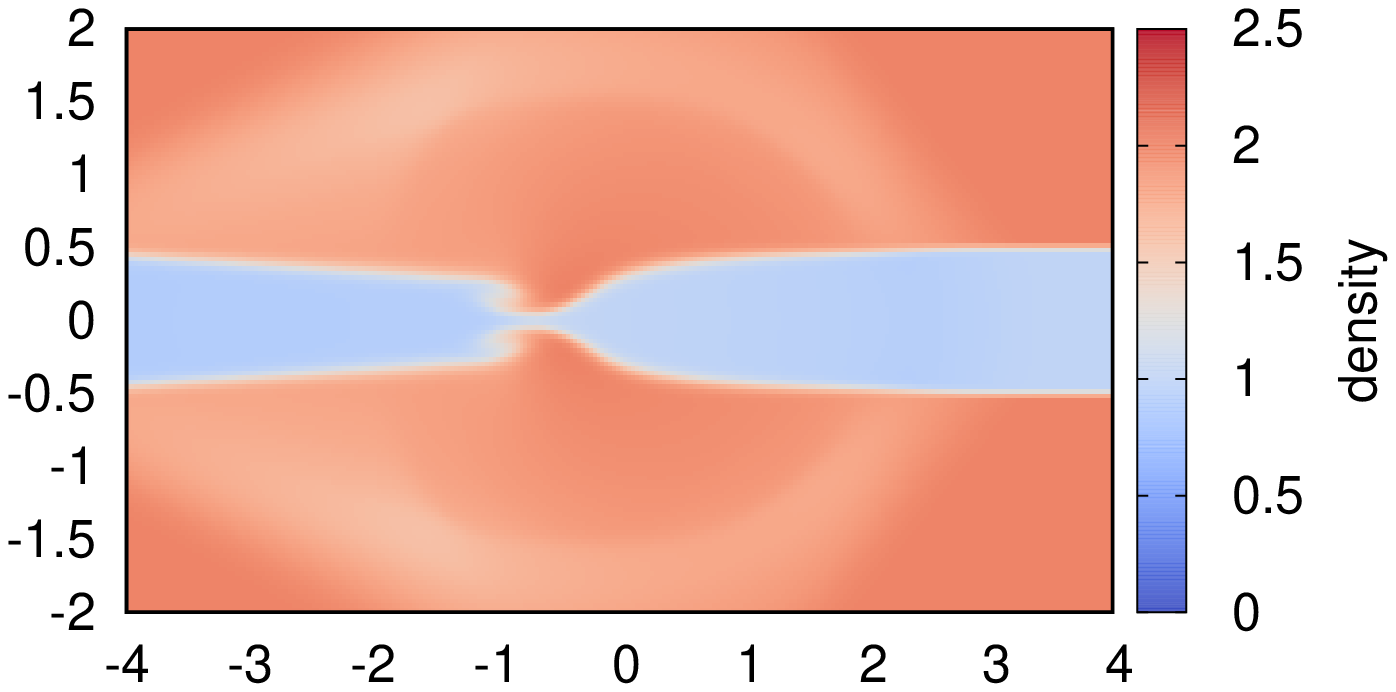}
         \caption{density at $t = 2.5$}
         \label{fig:result-plot-gas-d-t2-5}
      \end{subfigure}
      \hfill
   }
   \centerline{
      \begin{subfigure}[b]{0.45\textwidth}
         \includegraphics[width=\textwidth]{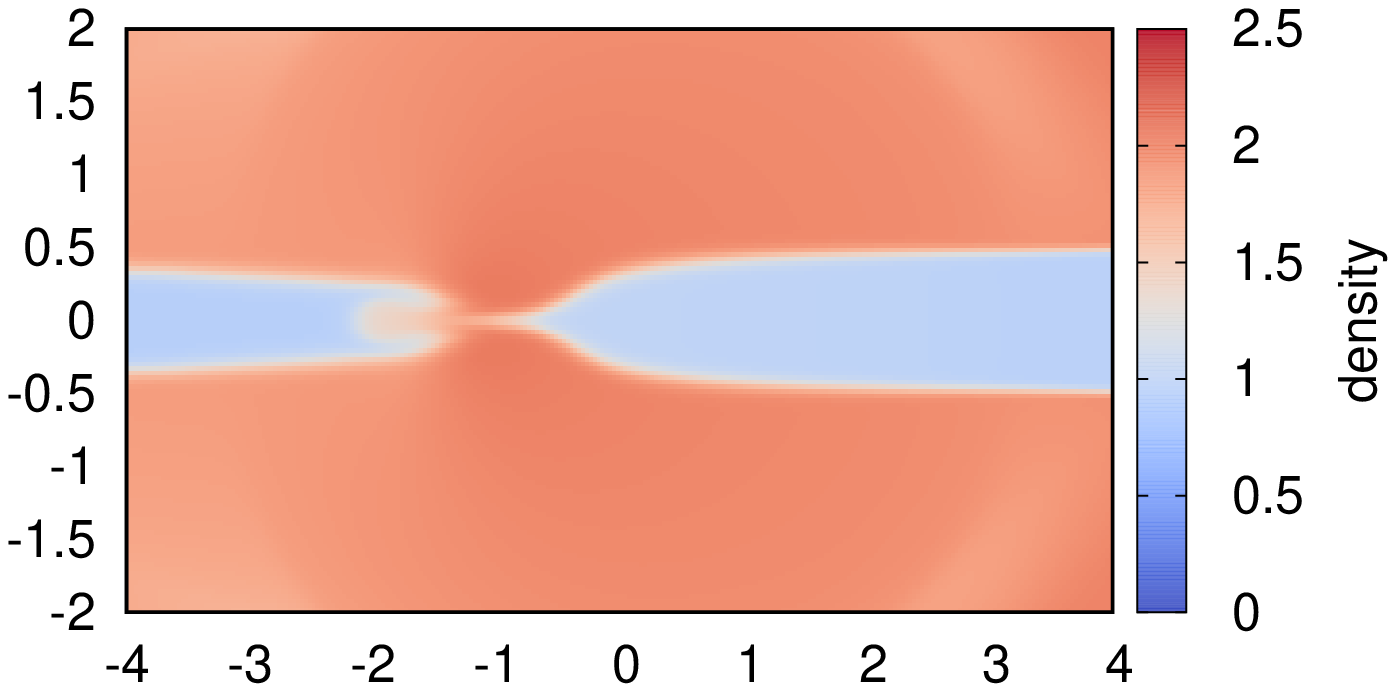}
         \caption{density at $t = 4.0$}
         \label{fig:result-plot-gas-d-t4-0}
      \end{subfigure}
   }
   \caption{Plots of density evolution for run B ($\tau = 2.0$, $u_\mathrm{init} = -1.0$)
            of the 2D gas simulation with a grid resolution of 32 points per unit length.
            After an initial set-up period, the thinning does not propagate tailward.}
   \label{fig:result-plot-gas}
\end{figure}

In the previous section,
of the three plasma parameters defined earlier,
we used only the initial velocity, $u_\mathrm{init}$.
We now extend the non-magnetized plasma model to 2D
by adding (so far unmagnetized) north and south ``lobe'' regions.
This introduces the second parameter, sheet/lobe temperature ratio $\tau$.
Since the lobe magnetic field is zero,
the pressure balance is achieved by
making the initial pressure uniform over the entire simulation area.

The initial conditions for simulation runs A, B, and C are shown in table~{\ref{tab:initial-gas}},
where variables $X_\mathrm{sheet}$ are for the plasma sheet ($\vect{U}_\mathrm{L}$, $\vect{U}_\mathrm{R}$),
and variables $X_\mathrm{lobe}$ are for the magnetic lobes ($\vect{U}_\mathrm{D}$, $\vect{U}_\mathrm{U}$).
The magnetic fields are all set to zero, $\vect{B}_\mathrm{sheet} = \vect{B}_\mathrm{lobe} = \vect{0}$.
Each of these configurations was simulated with
initial disturbance $\vect{U}_\mathrm{L}$, with velocity $u_\mathrm{init} = -1.0$.

2D plots of density for run B ($\tau = 2.0$, $u_\mathrm{init} = -1.0$)
are given in Fig.~{\ref{fig:result-plot-gas}}.
At time $t = 0$, the left half of the plasma sheet ($x < 0$, $-0.5 < y < 0.5$)
begins moving to the left at velocity $u_\mathrm{init}$.
This creates a drop in pressure in the centre of the plasma sheet,
which starts pulling in the surrounding plasma.
As a result, a rarefaction wave starts spreading in all directions.
The plasma from the right half of the sheet is pulled by the rarefaction wave,
lowering pressure and breaking the balance between sheet and lobes.

As the pressure balance is disturbed,
lobe plasma starts pushing at the sheet plasma,
transforming the rarefaction wave into plasma sheet thinning
(Fig.~{\ref{fig:result-plot-gas}}({\subref{fig:result-plot-gas-d-t1-0}})).
However, due to the inward movement of the lobe plasma
the pressure balance between sheet and lobes is quickly re-established.
While the rarefaction wave itself continues to propagate in all directions,
since the jump between sheet and lobe pressures is lost,
there is no further significant inward movement of
the sheet--lobe boundary on the right-hand side ($x > 0$) after $t \gtrsim 2.5$
(Figs.~{\ref{fig:result-plot-gas}}({\subref{fig:result-plot-gas-d-t2-5}})
and~({\subref{fig:result-plot-gas-d-t4-0}})).

%%%%%%%%%%%%%%%%%%%%%%%%%%%%%%%%%%%%%%%%%%%%%%%%%%%%%%%%%%%%%%%%%%%%%%%%%%%%%%%%
%%%%%%%%%%%%%%%%%%%%%%%%%%%%%%%%%%%%%%%%%%%%%%%%%%%%%%%%%%%%%%%%%%%%%%%%%%%%%%%%

\section{Plasma model, simulation and results}
\label{sec:results-plasma}

\begin{table}
   \centering
   \begin{tabular}{c||c|c|c||c|c|c|c|c||c|c|c|}
        Run
      & $\tau$
      & $\beta_\mathrm{lobe}$
      & $u_\mathrm{init}$
      & $\rho_\mathrm{sheet}$
      & $p_\mathrm{sheet}$
      & $\rho_\mathrm{lobe}$
      & $p_\mathrm{lobe}$
      & $B_{x,\mathrm{lobe}}$
      & $c_\mathrm{s,sheet}$
      & $c_\mathrm{s,lobe}$
      & $c_\mathrm{A,lobe}$ \\
      \hline
      D1 & $1.0$ & $1.0$ & $-1.0$ & $1.0$ & $1.0$ & $0.50$ & $0.50$ & $1.0$ & $1.29$ & $1.29$ & $1.41$ \\
      D2 & $1.0$ & $0.2$ & $-1.0$ & $1.0$ & $1.0$ & $0.16$ & $0.16$ & $1.3$ & $1.29$ & $1.29$ & $3.30$ \\
      \hline
      E1 & $2.0$ & $1.0$ & $-1.0$ & $1.0$ & $1.0$ & $1.00$ & $0.50$ & $1.0$ & $1.29$ & $0.91$ & $1.00$ \\
      E2 & $2.0$ & $0.2$ & $-1.0$ & $1.0$ & $1.0$ & $0.31$ & $0.16$ & $1.3$ & $1.29$ & $0.91$ & $2.33$ \\
      \hline
      F1 & $5.0$ & $1.0$ & $-1.0$ & $1.0$ & $1.0$ & $2.50$ & $0.50$ & $1.0$ & $1.29$ & $0.58$ & $0.63$ \\
      F2 & $5.0$ & $0.2$ & $-1.0$ & $1.0$ & $1.0$ & $0.78$ & $0.16$ & $1.3$ & $1.29$ & $0.58$ & $1.48$ \\
      \hline
      \hline
      ideal & $10.0$ & $0.002$ & --- & $1.0$ & $1.0$ & $0.02$ & $0.002$ & $1.413$ & $1.29$ & $0.40$ & $10.0$ \\
      \hline
   \end{tabular}
   \caption{An overview of the initial conditions for 2D plasma sheet simulations.
            The last row shows the ideal, realistic values, which couldn't be used due to limitations of the simulation program.}
   \label{tab:initial-mhd}
\end{table}

\begin{figure}
   \centerline{
      \hfill
      \begin{subfigure}[b]{0.45\textwidth}
         \includegraphics[width=\textwidth]{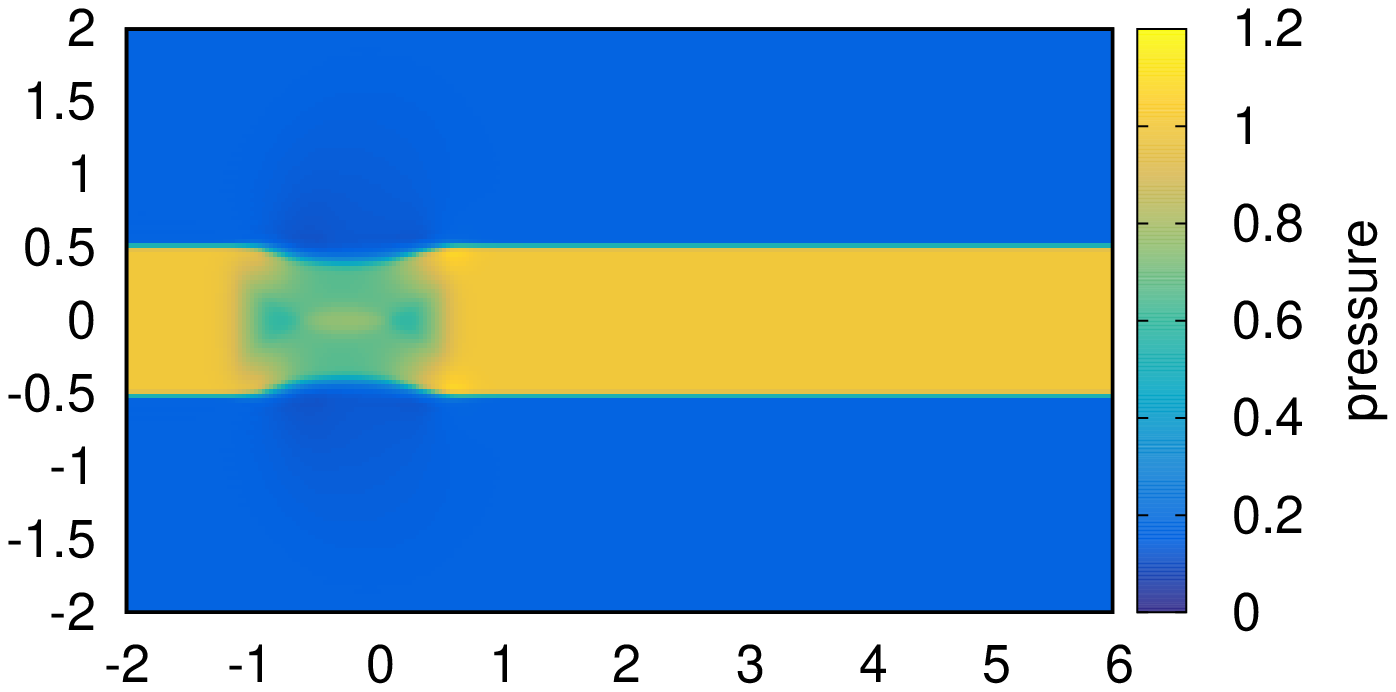}
         \caption{pressure at $t = 0.5$}
         \label{fig:result-plot-plasma-p-t0-5}
      \end{subfigure}
      \hfill
      \begin{subfigure}[b]{0.45\textwidth}
         \includegraphics[width=\textwidth]{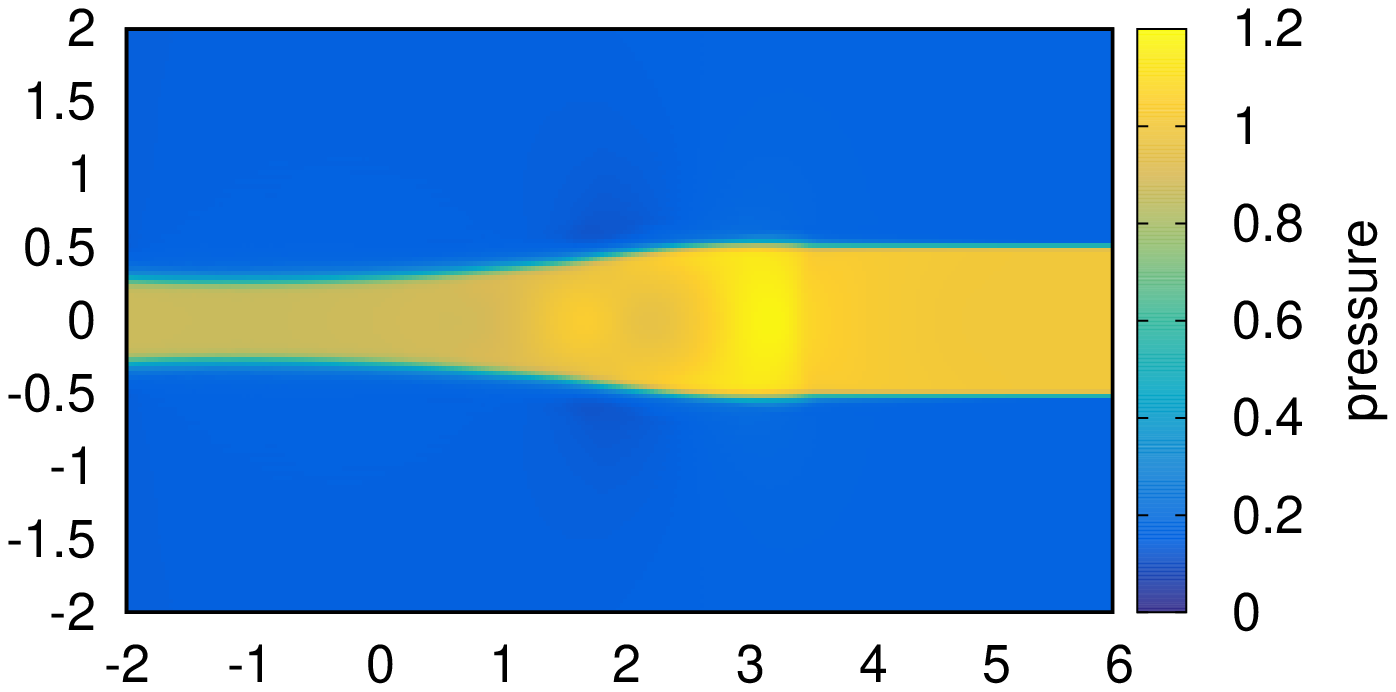}
         \caption{pressure at $t = 2.5$}
         \label{fig:result-plot-plasma-p-t2-5}
      \end{subfigure}
      \hfill
   }
   \centerline{
      \begin{subfigure}[b]{0.45\textwidth}
         \includegraphics[width=\textwidth]{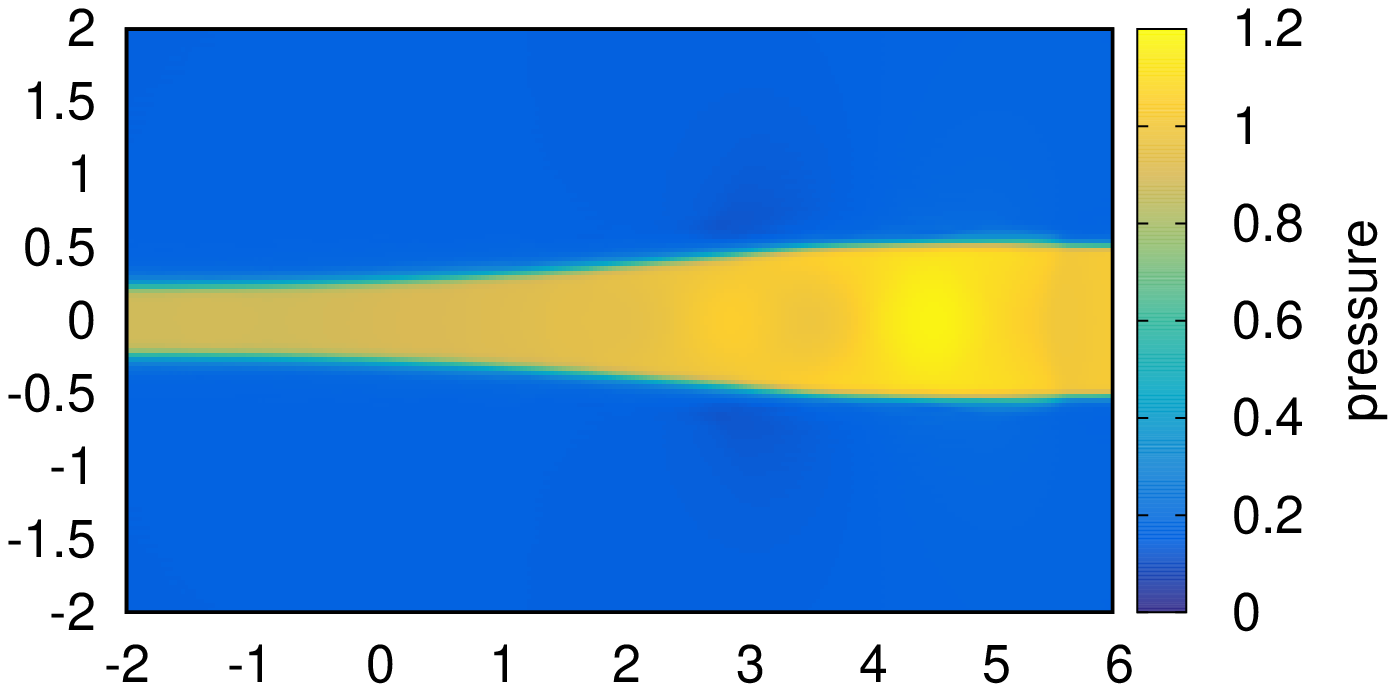}
         \caption{pressure at $t = 4.0$}
         \label{fig:result-plot-plasma-p-t4-0}
      \end{subfigure}
   }
   \caption{Plots of pressure evolution for run E2
            ($\tau = 2.0$, $\beta_\mathrm{lobe} = 0.2$, $u_\mathrm{init} = -1.0$)
            of the 2D plasma simulation with a resolution of 32 grid points per unit length.
            ({\subref{fig:result-plot-plasma-p-t0-5}})
            shows the initial plasma sheet pressure drop and the beginning of recompression.
            ({\subref{fig:result-plot-plasma-p-t2-5}}) and~({\subref{fig:result-plot-plasma-p-t4-0}})
            show the propagation of the thinning front.}
   \label{fig:result-plot-mhd}
\end{figure}

\begin{figure}
   \centerline{
      \hfill
      \begin{subfigure}[b]{0.45\textwidth}
         \includegraphics[width=\textwidth]{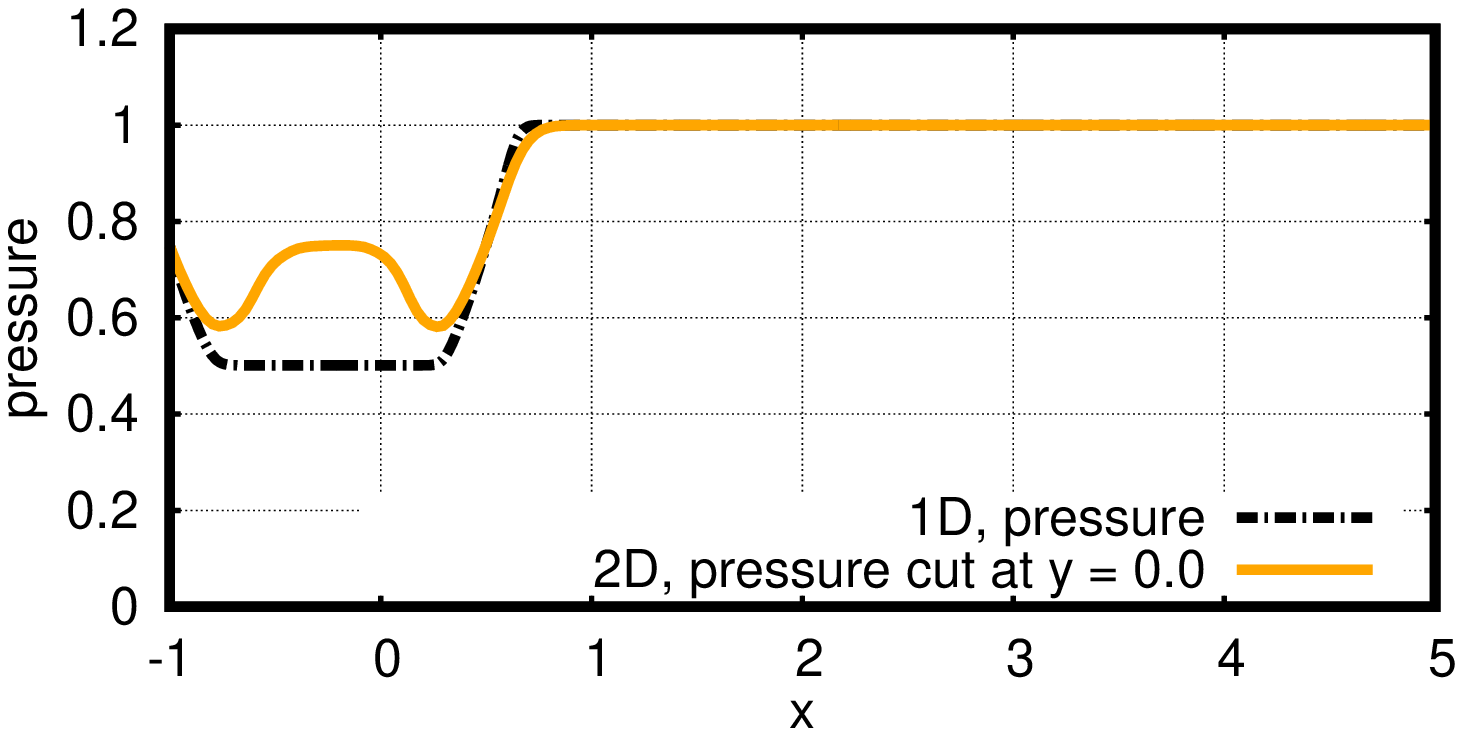}
         \caption{1D vs 2D at $t = 0.5$}
         \label{fig:result-mhd-cut-t05}
      \end{subfigure}
      \hfill
      \begin{subfigure}[b]{0.45\textwidth}
         \includegraphics[width=\textwidth]{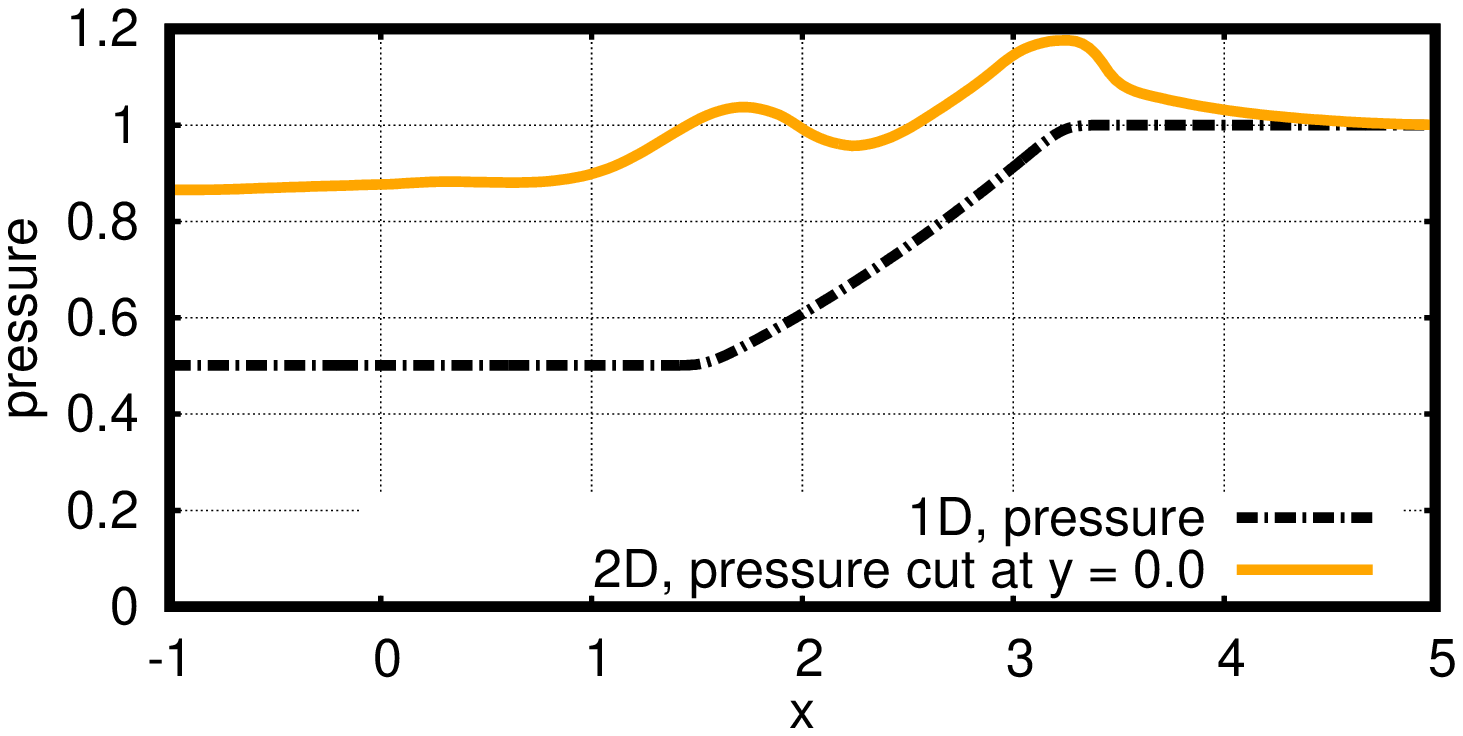}
         \caption{1D vs 2D at $t = 2.5$}
         \label{fig:result-mhd-cut-t25}
      \end{subfigure}
      \hfill
   }
   \centerline{
      \begin{subfigure}[b]{0.45\textwidth}
         \includegraphics[width=\textwidth]{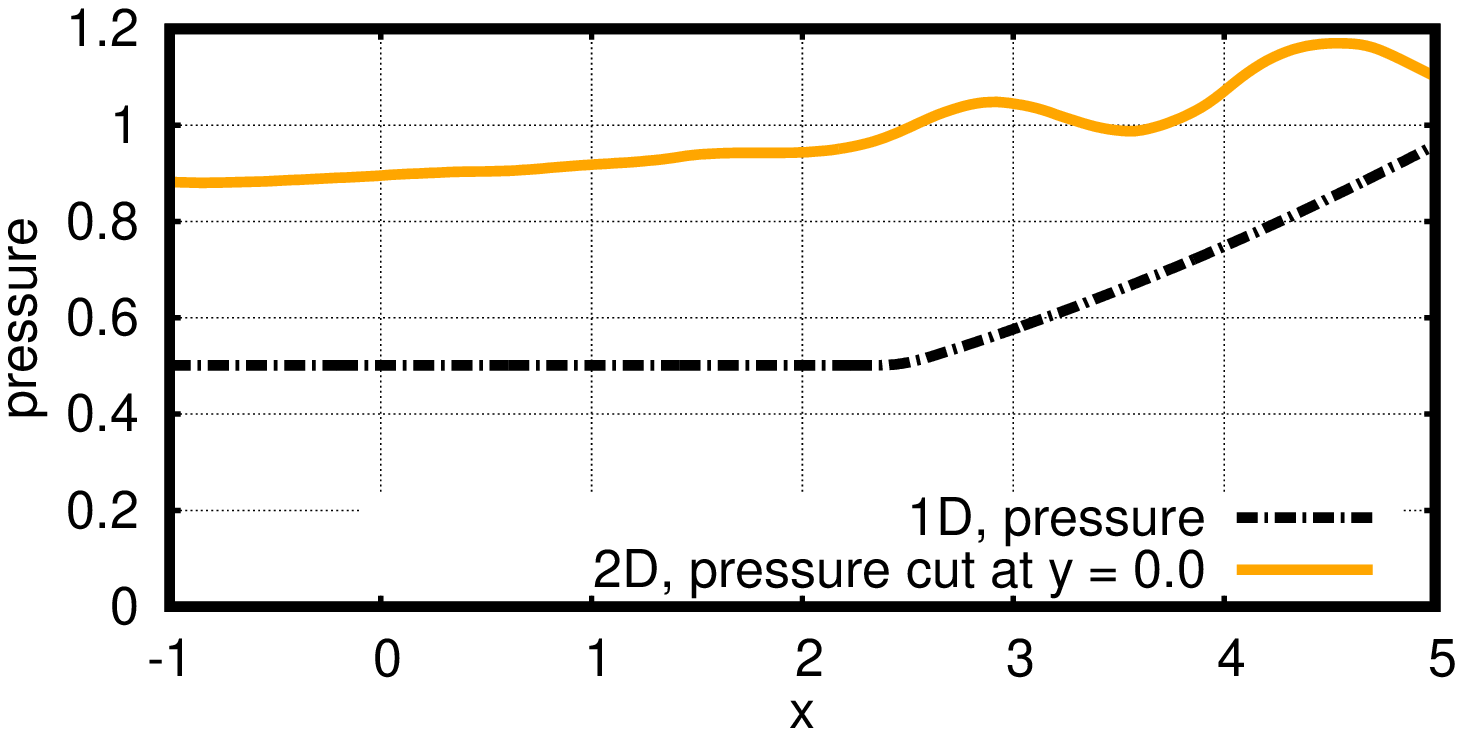}
         \caption{1D vs 2D at $t = 4.0$}
         \label{fig:result-mhd-cut-t40}
      \end{subfigure}
   }
   \caption{Comparison of pressure in 1D and 2D simulation for run E2
            ($\tau = 2.0$, $u_\mathrm{init} = -1.0$, $\beta_\mathrm{lobe} = 0.2$).
            The grid densities are, respectively, 128 and 32 points per unit length.
            For the 2D simulation, we show the horizontal cut through the centre of the plasma sheet, at $y = 0$.}
   \label{fig:result-mhd-cut}
\end{figure}

\begin{figure}
   \centerline{
      \hfill
      \begin{subfigure}[b]{0.45\textwidth}
         \includegraphics[width=\textwidth]{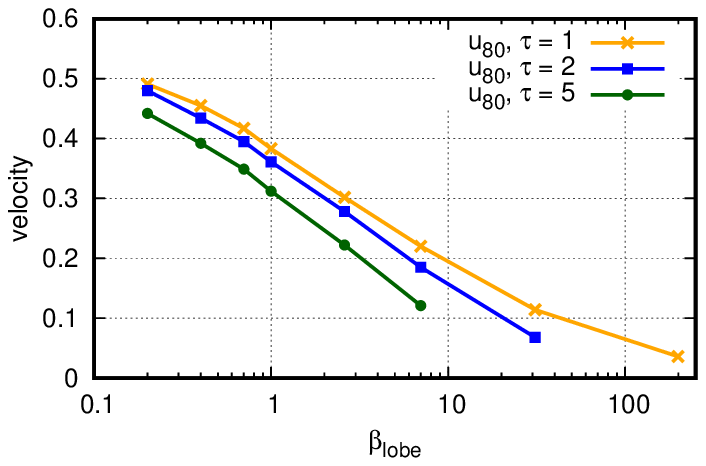}
         \caption{thinning velocity vs plasma beta}
         \label{fig:result-thin-vel-vs-beta}
      \end{subfigure}
      \hfill
      \begin{subfigure}[b]{0.45\textwidth}
         \includegraphics[width=\textwidth]{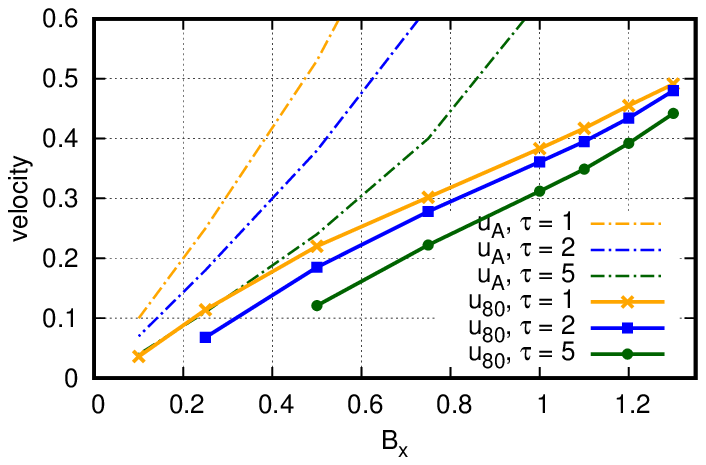}
         \caption{thinning velocity vs $B_x$}
         \label{fig:result-thin-vel-vs-bx}
      \end{subfigure}
      \hfill
   }
   \caption{Front velocity dependence on the temperature ratio $\tau$ and the plasma beta $\beta_\mathrm{lobe}$ (left)
            for 2D simulations with grid density of 32 points,
            plotted versus lobe beta (left) and the initial lobe magnetic field strength (right)
            with Alfv\'en velocities for the lobe initial conditions shown for comparison.}
   \label{fig:result-thin-vel}
\end{figure}

\begin{figure}
   \centerline{
      \hfill
      \begin{subfigure}[b]{0.45\textwidth}
         \includegraphics[width=\textwidth]{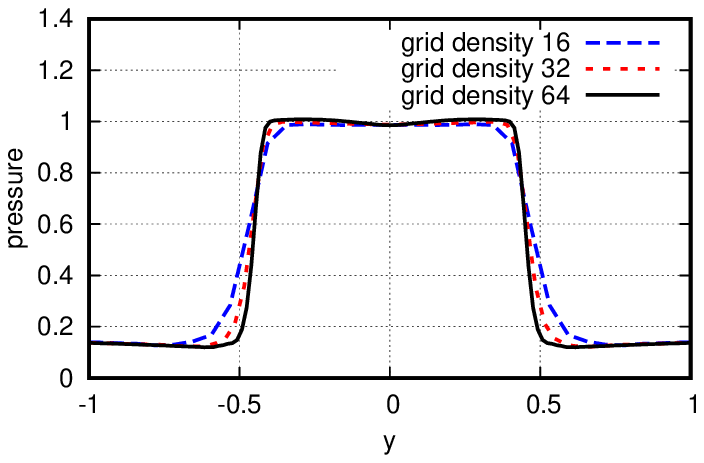}
         \caption{pressure profiles for run E2 at $x = 2.0$}
         \label{fig:grid-density-comparison-y-profile}
      \end{subfigure}
      \hfill
      \begin{subfigure}[b]{0.45\textwidth}
         \includegraphics[width=\textwidth]{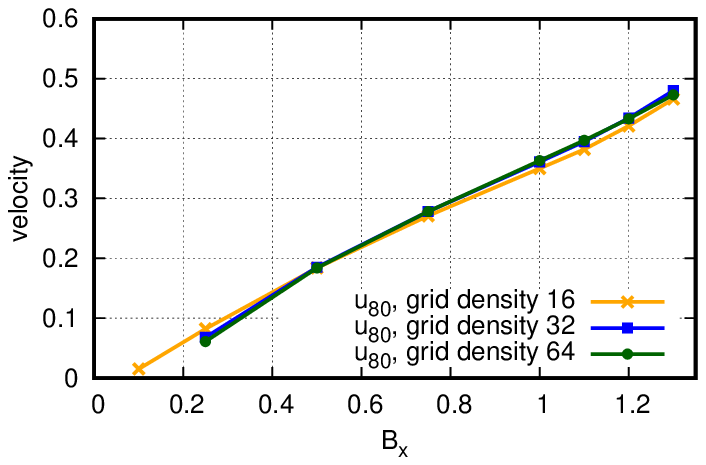}
         \caption{thinning velocity vs $B_x$}
         \label{fig:grid-density-comparison-thinning-velocity}
      \end{subfigure}
      \hfill
   }
   \caption{Comparisons between simulation results with different grid densities.
            Pressure profiles of the plasma sheet are increasingly smeared out as grid density falls (left),
            though the sheet width is consistent.
            The impact of the minor width variation on measured thinning velocity is low (right).}
   \label{fig:grid-density-comparison}
\end{figure}

Finally, we introduce the lobe magnetic field, described with the third parameter,
lobe plasma beta $\beta_\mathrm{lobe}$.
A realistic lobe plasma beta would be on the order of
$\beta_\mathrm{lobe} \lesssim 0.01$~{\citep{baumjohann1990pressure}};
however, this is difficult to achieve in a simulation
due to the extremely low kinetic pressure in such a plasma.
For this paper, we limit the values of plasma beta to $\beta_\mathrm{lobe} \geq 0.2$.

The addition of the magnetic field to the lobe plasma means that,
to keep the total pressure constant and the lobe/sheet pressure balanced,
the lobe kinetic pressure must be lowered.
An overview of the initial conditions is shown in table~{\ref{tab:initial-mhd}}.

2D plots of pressure for run E2 are shown in Fig.~{\ref{fig:result-plot-mhd}}.
At time $t = 0$, the left half of the plasma sheet begins moving earthward.
The pressure drop that the disturbance leaves behind pulls in the surrounding plasma
(Fig.~{\ref{fig:result-plot-mhd}}({\subref{fig:result-plot-plasma-p-t0-5}})).
Figure~{\ref{fig:result-mhd-cut}} shows the time evolution of sheet pressure
taken from 1D gas simulation at a grid density of 128 points (black dash-dotted line)
and a profile at $y = 0$ taken from the 2D plasma simulation (orange solid line).
For a few moments, the resulting rarefaction wave in 2D is similar
to the one in the 1D simulation ($0.2 \lesssim x \lesssim 0.8$ in
Fig.~{\ref{fig:result-mhd-cut}}({\subref{fig:result-mhd-cut-t05}}));
however, as the boundaries with the magnetic lobes move inward due to loss of the pressure balance,
the plasma sheet is compressed and the rarefaction wave is no longer visible
(Figs.~{\ref{fig:result-mhd-cut}}({\subref{fig:result-mhd-cut-t25}})
and~({\subref{fig:result-mhd-cut-t40}})).
Despite the apparent loss of the rarefaction wave,
the earthward plasma flow and the accompanying thinning of the plasma sheet continue.
It is noted that the thinning propagates in a self-similar fashion
after an initial set-up period at $t \lesssim 1$
(see Figs.~{\ref{fig:result-plot-mhd}}({\subref{fig:result-plot-plasma-p-t2-5}})
and~({\subref{fig:result-plot-gas-d-t4-0}})).

It is worth noting that we observed the development
of Kelvin-Helmholtz instability on the sheet--lobe interface,
arising due to the velocity difference between the two plasmas~{\citep{chandrasekhar1981book-kh}}.
However, the instability appears only for the weak magnetic field
($B_{x, \mathrm{lobe}} \lesssim 0.5$, $\beta_\mathrm{lobe} \gtrsim 7$),
and when it does appear its effect is constrained
to the far left of the simulation domain ($x \lesssim -3$),
where the velocity difference is significantly larger.
As we are only interested in the right side of the domain ($x > 0$),
presence of the instability does not affect the following discussions.

For further analysis of the plasma sheet thinning,
we measure the propagation velocity of its front with the following method.
First, for each discrete $x$ coordinate, linearly interpolate the $B_x$ profile
to find the location in $y$ where the magnetic field drops below half of the initial lobe value.
Collecting all the $(x,y)$ values gives us a rough profile of the plasma sheet,
which we can again linearly interpolate to obtain the $x$ coordinate of
the point where the sheet thickness is reduced to $80\%$ of the initial value.
We repeat the procedure and obtain the ``$80\%$ thinning'' locations between $t = 2$ and $t = 10$
(skipping the initial period at $t < 2.0$ where the self-similar shape of the thinning may not yet be fully developed).
Finally, we derive the thinning velocity $u_{80}$ from a linear fit on the ``$80\%$ thinning'' locations.
The results for grid density $32$ are shown in Fig.~{\ref{fig:result-thin-vel}}.

We can observe a strong, approximately linear dependence between
the magnetic field strength in the lobes and the thinning velocity.
Furthermore, all thinning velocities are considerably slower than the sheet sound velocity.
Assuming the linear relationship holds for smaller $\beta_\mathrm{lobe}$ (larger $B_{x,\mathrm{lobe}}$),
and taking into the account that $B_{x,\mathrm{lobe}} \leq \sqrt{2}$,
we can extrapolate that the maximum velocity of the $80\%$ thinning is $u_{80} \sim 0.65$,
or about half of the sheet sound velocity $c_\mathrm{s,sheet} = 1.29$.

In order to show numerical convergence,
the comparison between grid densities for temperature ratio $\tau = 2$
is shown in Fig.~{\ref{fig:grid-density-comparison}}.
We can see a good agreement between grid densities $16$ and $32$,
and an excellent agreement between grid densities $32$ and $64$.
More specifically, for $\beta_\mathrm{lobe} \leq 7$ ($B_{x,\mathrm{lobe}} > 0.5$),
the disagreement in thinning velocity between grid densities $16$ and $32$ is below $4\%$,
and the disagreement between grid densities $32$ and $64$ is below $2\%$.
The convergence worsens for $\beta_\mathrm{lobe} > 7$,
as determining the thinning velocity grows less reliable
and requires more grid density as that velocity nears zero.
However, lobe plasma is a low-beta plasma, and
the simulations for $\beta_\mathrm{lobe} > 1$ are used only to
confirm that the thinning velocity goes to zero as lobe beta rises
(as anticipated from the 2D gas simulation);
therefore, somewhat rough results for high values of lobe beta are acceptable.

%%%%%%%%%%%%%%%%%%%%%%%%%%%%%%%%%%%%%%%%%%%%%%%%%%%%%%%%%%%%%%%%%%%%%%%%%%%%%%%%
%%%%%%%%%%%%%%%%%%%%%%%%%%%%%%%%%%%%%%%%%%%%%%%%%%%%%%%%%%%%%%%%%%%%%%%%%%%%%%%%

\section{Comparison of 1D and 2D models}
\label{sec:comparison}

In the 1D gas model~{\citep{chao_plasma}},
the initial disturbance generates a rarefaction wave travelling tailward.
As the pressure drops behind the wave,
the boundary between sheet and lobe is forced to move inward.
Since the plasma beta in the sheet is greater than one,
the rarefaction wave moves at the sound velocity, $c_\mathrm{s,sheet}$.
The thinning front, presumably, moves at approximately the same velocity.
There is no stated dependence between the thinning front velocity
and the conditions in the magnetic lobes;
only the thinning amount is influenced by them.

Extending to a 2D gas model, we observed that the rarefaction wave
generated by the initial disturbance
is drastically weakened in the first moments of the event.
Furthermore, even though this weakened form of
the rarefaction wave continues propagating,
the thinning ceases to propagate due to
the loss of the sheet--lobe pressure difference (Fig.~{\ref{fig:result-plot-gas}}).

Extending to a 2D plasma model by
introducing the magnetic field into the lobes,
the dynamics of the plasma sheet thinning drastically changes.

First, the rarefaction wave is weakened so much
that it is no longer clearly noticeable as an independent entity.
It is either subsumed in other, stronger waves,
or completely extinguished by compression in the first few moments of the event.
In either case, the significant drop in pressure
that should have been what causes the thinning to propagate is absent.
However, despite the lack of a significant pressure drop,
and in stark contrast to the 2D gas model,
the thinning continues to propagate (Fig.~{\ref{fig:result-plot-mhd}}).
This indicates that the rarefaction wave
is not a sole component of the plasma sheet thinning.

Second, the thinning front velocity is lower than the rarefaction wave velocity.
This is another indicator that thinning dynamics
have separated from the rarefaction wave that initially caused them.

Third, the thinning front velocity shows a strong dependence
on the conditions in the magnetic lobes (Fig.~{\ref{fig:result-thin-vel}}).
The thinning front propagates faster when the lobe magnetic field is stronger,
in stark contrast to the 1D model.
As the 2D gas simulation showed, in the limit of no magnetic field there is an initial burst
after which the thinning front completely stops propagating;
this aligns with the small $B_x$ limit of the 2D plasma simulation.

The above points of comparison clearly show that the dynamics of plasma sheet thinning
seem to be dominated by the sheet--lobe interaction that could not be accounted for in the 1D model.

Finally, as mentioned in Section~{\ref{sec:introduction}},
the CD model employs the rarefaction wave and the resulting pressure drop
to trigger reconnection in the near Earth neutral line.
That is, the rarefaction wave needs to propagate tailward for a long distance,
for example, from the site of current disruption ($\sim -10 R_\textrm{e}$)
to that of reconnection ($\sim -20 R_\textrm{e}$), which are approximately $10 R_\textrm{e}$ apart.
In the current simulation study, however,
the rarefaction wave and, more importantly, corresponding pressure drop
are damped soon after the occurrence of pressure decrease due to current disruption.
This would not allow reconnection to occur because of the absence of pressure decrease
in the central plasma sheet at $\sim -20 R_\textrm{e}$.
In this sense, the results of current numerical simulation
do not support the CD model for explaining all the phenomena
in the magnetotail during the episode of auroral substorm.

%%%%%%%%%%%%%%%%%%%%%%%%%%%%%%%%%%%%%%%%%%%%%%%%%%%%%%%%%%%%%%%%%%%%%%%%%%%%%%%%
%%%%%%%%%%%%%%%%%%%%%%%%%%%%%%%%%%%%%%%%%%%%%%%%%%%%%%%%%%%%%%%%%%%%%%%%%%%%%%%%

\section{Conclusion}
\label{sec:conclusion}

Starting from a simple 1D model of the plasma sheet thinning,
we have first extended it to a 2D configuration
by adding the north and south (non-magnetic) lobes.
In this 2D gas simulation the rarefaction wave is weakened
and thinning ceases to propagate (Fig.~{\ref{fig:result-plot-gas}}).
After adding a magnetic field to the lobes and simulating the resulting 2D plasma model,
we observed that the thinning propagates again,
but this time the rarefaction wave is absent (Fig.~{\ref{fig:result-plot-mhd}}).

The lack of thinning propagation in the 2D gas simulation
means that the influence of the sheet--lobe configuration
on the dynamics can be too strong to allow extrapolating the behaviour from a 1D model.
The appearance of thinning in the 2D plasma simulation
indicates that the deformation of the magnetic field may play a significant role
in the plasma sheet thinning.
This conclusion is strengthened by observing that
the signature aspect of the Current Disruption model
of the plasma breakup, the rarefaction wave,
as well as its associated pressure drop, are drastically weakened soon after the event begins,
and the thinning, which was supposed to be following behind the---now absent---rarefaction wave,
continues propagating, though at a slower velocity.
The thinning velocity is shown to be strongly influenced by the conditions in the magnetic lobes;
in particular, there is an approximately linear dependence on the lobe magnetic field strength.

Finally, the weakening or outright disappearance of the rarefaction wave,
as well as the shape of the plasma sheet thinning,
indicate that the CD model may not be able to trigger magnetic reconnection.

In future research, we hope to more precisely determine
the dependence of the plasma sheet thinning
on the parameters of the plasma sheet and magnetic lobes.

%%%%%%%%%%%%%%%%%%%%%%%%%%%%%%%%%%%%%%%%%%%%%%%%%%%%%%%%%%%%%%%%%%%%%%%%%%%%%%%%
%%%%%%%%%%%%%%%%%%%%%%%%%%%%%%%%%%%%%%%%%%%%%%%%%%%%%%%%%%%%%%%%%%%%%%%%%%%%%%%%

%\section*{Acknowledgements}
%\label{sec:acknowledgements}

This research was partially supported by
the Rotary Yoneyama Memorial Scholarship.

%%%%%%%%%%%%%%%%%%%%%%%%%%%%%%%%%%%%%%%%%%%%%%%%%%%%%%%%%%%%%%%%%%%%%%%%%%%%%%%%
%%%%%%%%%%%%%%%%%%%%%%%%%%%%%%%%%%%%%%%%%%%%%%%%%%%%%%%%%%%%%%%%%%%%%%%%%%%%%%%%

\bibliographystyle{jpp}
\bibliography{citations}

\end{document}